\documentclass[usegraphicx,usenatbib,fleqn,useAMS]{mn2e}

\usepackage{amsmath}
\usepackage[varg]{txfonts}

\renewcommand\pi\upi
\newcommand\dm{\mathrm d}
\newcommand\msun{\mbox{M}_{\sun}}
\newcommand\alphap{{\tilde\alpha}}
\newcommand\muasa{{$\umu$as~a$^{-1}$}}

\newcommand\original{13 October 2011}
\newcommand\p{\phantom{0}}

\defcitealias{WEA}{WEA}
\defcitealias{AE11}{Paper~I}
\defcitealias{Ev11}{EAD}

\title[Mass Estimators in the Gaia era]
{Mass Estimators in the \emph{Gaia} era}

\author[An, Evans \& Deason]
{J.~An$^1$, N.~W.~Evans$^2$, and A.~J.~Deason$^2$
\\$^1$National Astronomical Observatories,
Chinese Academy of Sciences, A20 Datun Road, Chaoyang District,
Beijing 100012, PR~China;\thanks{E-mail:~jinan@nao.cas.cn}
\\$^2$Institute of Astronomy, University of Cambridge, Madingley Road,
Cambridge CB3~0HA.\thanks{E-mail:~nwe,ajd75@ast.cam.ac.uk}}

\date{Accepted. Received \today; in original form \original}
\pagerange{\pageref{firstpage}--\pageref{lastpage}}
\volume{000}\pubyear{2011}

\begin{document}
\label{firstpage}
\maketitle

\begin{abstract}\noindent
  Forthcoming astrometric missions such as the {\it Gaia} satellite
  will bring to the fore the problem of estimating the enclosed mass
  from a set of positions, radial velocities and proper motions of
  tracer stars. Here, we show how to construct the tracer mass
  estimator when the proper motion data are available in addition to
  the usual line-of-sight velocity data. Notably, the mass estimators
  do not require any assumption on the anisotropy, as it is naturally
  incorporated through the different components of proper motions. In
  addition, the separate treatment of the proper motions and the
  line-of-sight velocities is desirable because they are
  observationally independent and thus the propagation of the combined
  uncertainties is rather straightforward.  The extension to projected
  data is also sketched, together with a possible application of
  measuring the masses of Galactic globular clusters.
\end{abstract}

\section{Introduction}

The great era of Galactic Cartography is upon us. The last twenty
years have seen ambitious photometric and spectroscopic surveys of the
Galaxy provide magnitudes and radial velocities for many millions
of stars. Astrometric surveys of the Galaxy began with the now
comparatively modest {\it Hipparcos} mission (see \citealt{Pe09} for a
summary of its achievements) and will culminate with the launch of the
{\it Gaia} satellite in 2013. This will provide micro-arcsecond
astrometry for $\sim10^9$ stars brighter than $G\approx20$
\citep{Pe01,Tu05}, as well as radial velocities for stars brighter than
$G\approx17$. Positional and kinematic data on bright stars in the
Galaxy and its neighbours will be routinely available by the end of
its mission.

This wealth of new data motivates the study of new theoretical
questions. A classical problem in dynamical astronomy is to estimate
the enclosed mass from the relative positions and velocities of a
collection of bound test particles or tracers. Hitherto, work has
primarily concentrated on the case in which the velocities are
line-of-sight or projected. Here there has been an extensive body of
work starting with methods essentially exploiting the virial theorem
\citep{Pa52,Li60}. The problems with such virial estimators were
enunciated by \citet{BT81}, who introduced the notion of projected mass
estimators based on sums of the projected mass contributions,
namely [(projected distance) $\times$ (radial velocity)$^2$ / G].
In this form, the estimator is naturally suited to
the case of a point mass (the Kepler potential). The extension to
scale-free potentials with $\psi\propto r^{-\alpha}$ was
developed by \citet{Wh81}, \citet{Ev03} and \citet[hereafter WEA]{WEA}.
The estimator then takes a different form, namely sums of contributions
of the form [(distance)$^\alpha$ $\times$ (radial velocity)$^2$ / G].
Finally, \citet[hereafter Paper I]{AE11} have shown how to adapt these
ideas for an arbitrary spherical potential by choosing a weight function.

The case of proper motions has so far received very little
attention. The coming of age of {\it Gaia} will make the availability
of proper motions and radial velocities standard for all objects in
the nearby Universe brighter than $G\approx20$. Hence, this paper is
devoted to the problem of estimating the enclosed mass from a tracer
population when their space velocities are known.

%\bigskip
\section{Tracer mass estimators}

Throughout this paper, we operate under the simplest assumption that
the gravitational potential is spherical. A tracer
population of stars with a number density $\nu$ resides in the
potential. This is sampled to provide $N_\mathrm{tot}$ objects with
known positions, radial velocities and proper motions. We shall
usually consider the application to the problem of estimating the dark
halo mass within the radius of the outermost datapoint
$r_\mathrm{out}$.

We start with the spherical Jeans equation, which reads
\begin{equation}\label{eq:je}
\frac{\dm(\nu\sigma_r^2)}{\dm r}
+2\beta\frac{\nu\sigma_r^2}r
%+(2\sigma_r^2-\sigma_\mathrm t^2)\frac\nu r
%=\frac1{r^2}\frac{\dm(r^2\nu\sigma_r^2)}{\dm r}
%-\frac{\nu\sigma_\mathrm t^2}r
=-\nu\frac{GM}{r^2},
\end{equation}
where
\begin{equation}\label{eq:beta}
\beta(r)=1-\frac{\sigma_\theta^2+\sigma_\phi^2}{2\sigma_r^2}
=1-\frac{\sigma_\mathrm t^2}{2\sigma_r^2},
\end{equation}
is the \citet{Bi80} anisotropy parameter.
This can be integrated into the form \citepalias[see][eq.~11]{AE11}
\begin{equation}\label{eq:ae11}
GM_\mathrm{out}
=\left\langle\left[3+\hat\alpha(r)-2\beta(r)\right]
\frac{rv_r^2}{\tilde\mu}\right\rangle
-3r_\mathrm{out}\varsigma^2,
\end{equation}
where the angled bracket represents the average over the whole tracer
population. Here $M_\mathrm{out}$ is the total halo mass within radius
$r_\mathrm{out}$, and
\begin{equation}\label{eq:mprof}
\tilde\mu(r)\equiv\frac{M(r)}{M_\mathrm{out}}
\qquad(0\le r\le r_\mathrm{out})
\end{equation}
is the halo mass profile $M(r)$ normalized by the total mass
such that $\tilde\mu(r_\mathrm{out})=1$, whereas
\begin{equation}
\hat\alpha(r)\equiv-\frac{\dm\log[r^{-1}M(r)]}{\dm\log r}
=1-\frac{\dm\log\tilde\mu(r)}{\dm\log r},
\end{equation}
is the effective index for the gravitational potential.
Finally, the surface term is given by
\begin{equation}
\varsigma^2
\equiv\frac{\nu(r_\mathrm{out})\,\sigma_r^2(r_\mathrm{out})}
{\bar\nu_\mathrm{out}}
\,;\qquad
\bar\nu_\mathrm{out}\equiv\frac{3N_\mathrm{tot}}{4\pi r_\mathrm{out}^3}.
\end{equation}
Note that $\bar\nu_\mathrm{out}$ is the mean number density of the
tracers in the sphere of the radius of $r_\mathrm{out}$.

Next, we observe from equation (\ref{eq:beta}) that
\begin{equation}
\left\langle\frac{rv_\mathrm t^2}{\tilde\mu}\right\rangle
=\frac{4\pi}{N_\mathrm{tot}}\!\int_0^{r_\mathrm{out}}\!
\frac{\nu\sigma_\mathrm t^2r^3}{\tilde\mu}\,\dm r
=\left\langle2(1-\beta)\frac{rv_r^2}{\tilde\mu}\right\rangle
\end{equation}
where $v_\mathrm t=(v_\theta^2+v_\phi^2)^\frac12$ is the tangential
velocity and $\sigma_\mathrm t=(\sigma_\theta^2+\sigma_\phi^2)^\frac12$
is the corresponding velocity dispersion. Hence,
equation (\ref{eq:ae11}) is equivalent to
\begin{equation}\label{eq:main}
GM_\mathrm{out}
=\left\langle\frac{(1+\hat\alpha)rv_r^2}{\tilde\mu}\right\rangle
+\left\langle\frac{rv_\mathrm t^2}{\tilde\mu}\right\rangle
-3r_\mathrm{out}\varsigma^2,
\end{equation}
which is our desired result expressing the mass $M_\mathrm{out}$ in
terms of averages of observables.

The same can also be derived by directly integrating
equation (\ref{eq:je}) whose left side reduces to
\begin{subequations}
\begin{equation}\label{eq:je0}
\frac{\dm(\nu\sigma_r^2)}{\dm r}
%+2\beta\frac{\nu\sigma_r^2}r
+(2\sigma_r^2-\sigma_\mathrm t^2)\frac\nu r
=\frac1{r^2}\frac{\dm(r^2\nu\sigma_r^2)}{\dm r}
-\frac{\nu\sigma_\mathrm t^2}r.
%=-\nu\frac{GM}{r^2}.
\end{equation}
With an arbitrary weight $h=h(r)$, this then results in
\begin{equation}\label{eq:jeint}
\int_0^{r_\mathrm{out}}\!\dm r\,
hr^2\frac{\dm(r^2\nu\sigma_r^2)}{\dm r}
-\int_0^{r_\mathrm{out}}\!\dm r\,h\nu\sigma_\mathrm t^2r^3
={-\int_0^{r_\mathrm{out}}\!\dm r\,h\nu GMr^2}.
\end{equation}
The first integral here reduces to
\begin{multline}\label{eq:intp}
\int_0^{r_\mathrm{out}}\!\dm r\,hr^2
\frac{\dm(r^2\nu\sigma_r^2)}{\dm r}
=hr^4\nu\sigma_r^2\bigr\rvert^{r_\mathrm{out}}_0
-\int_0^{r_\mathrm{out}}\!\dm r\,r^2\nu\sigma_r^2
\frac{\dm(hr^2)}{\dm r}
\\=hr^4\nu\sigma_r^2\bigr\rvert^{r_\mathrm{out}}_0
-\int_0^{r_\mathrm{out}}\!\dm r\,
\frac{\dm\log(hr^2)}{\dm\log r}\,h\nu\sigma_r^2r^3
\end{multline}
\end{subequations}
through integration by parts. Given that the inner boundary
term $hr^4\nu\sigma_r^2\rvert_0=0$ vanishes, replacing the first
integral in equation (\ref{eq:jeint}) by equation (\ref{eq:intp})
and multiplying the result by $4\pi N_\mathrm{tot}^{-1}$, we find that
\begin{equation}\label{eq:mm}
\left\langle hGM\right\rangle=
\left\langle\biggl(2+\frac{\dm\log h}{\dm\log r}\biggr)\,
hrv_r^2\right\rangle+\left\langle hrv_\mathrm t^2\right\rangle
-3r_\mathrm{out}\varsigma^2h(r_\mathrm{out}).
\end{equation}
We recover equation (\ref{eq:main}) by choosing $h=\tilde\mu^{-1}$
since $\dm\log h/\dm\log r={-\dm\log\tilde\mu/\dm\log r}=\hat\alpha-1$.

%\vfill
%\medskip
\subsection{Power-Law Weight}
\subsubsection{Scale-free potentials}

\citet[hereafter EAD]{Ev11} have demonstrated that for most practical
purposes adopting the scale-free assumption is a good approximation
for choosing the proper mass estimator. This amounts to assuming that
the mass model is given by
\begin{equation}\label{eq:mss}
\tilde\mu(r)
%=\frac{M(r)}{M_\mathrm{out}}
=\biggl(\frac r{r_\mathrm{out}}\biggr)^{1-\alpha}
\quad(0\le r\le r_\mathrm{out}),
\end{equation}
and so $\hat\alpha=\alpha$ is constant. The exact models of the Kepler
potential (a central point mass), the logarithmic potential (a
truncated singular isothermal sphere), and the harmonic potential (a
uniform homogeneous sphere) correspond to the cases that $\alpha=1$,
$\alpha=0$, and $\alpha=-2$, respectively. With the mass model of
equation (\ref{eq:mss}), equation (\ref{eq:main}) reduces to
\begin{equation}\label{eq:mep}
\frac{GM_\mathrm{out}}{r_\mathrm{out}^{1-\alpha}}
=(1+\alpha)\left\langle v_r^2r^\alpha\right\rangle
+\left\langle v_\mathrm t^2r^\alpha\right\rangle
-3r_\mathrm{out}^\alpha\varsigma^2.
\end{equation}
Given that the tracer density profile essentially
vanishes at finite $r_\mathrm{out}$,
the surface term should drop out
(unless they are pressure-confined with a hard cut-off).
Alternatively, following \citetalias{WEA},
the surface term may be explicitly solved under
the assumption of exact power-law behaviour for the
tracer density ($\nu\propto r^{-\gamma}$) and the velocity
dispersions, although strictly speaking the resulting solutions cannot
be physical everywhere. It follows that %the anisotropy parameter
$\beta$ is constant since the Jeans equation implies that
$\sigma_r^2$, $\sigma_\mathrm t^2$, and $GMr^{-1}$ all follow the same
power-law. Therefore we can solve for
$\nu(r_\mathrm{out})/\bar\nu_\mathrm{out}=\frac13(3-\gamma)$ and
$\sigma_r^2=(\gamma-2\beta+\alpha)^{-1}GMr^{-1}$. If we substitute
these for the surface term, then we find that:
\begin{equation}
\frac{GM_\mathrm{out}}{r_\mathrm{out}}
+3\varsigma^2
=\frac{3-2\beta+\alpha}{\gamma-2\beta+\alpha}
\frac{GM_\mathrm{out}}{r_\mathrm{out}}.
\end{equation}
Together with $\langle v_\mathrm t^2r^\alpha\rangle=2(1-\beta)
\langle v_r^2r^\alpha\rangle$, equation (\ref{eq:mep}) results in
\begin{equation}\label{eq:mpp}
\frac{GM_\mathrm{out}}{r_\mathrm{out}^{1-\alpha}}
=(\alpha+\gamma-2)\left\langle v_r^2r^\alpha\right\rangle
+\left\langle v_\mathrm t^2r^\alpha\right\rangle.
\end{equation}
This is equivalent to equations (16) and (24) of \citetalias{WEA}
if the anisotropy parameter is given by %the effective value
$\beta=1-\langle v_\mathrm t^2r^\alpha\rangle/(2\langle
v_r^2r^\alpha\rangle)$.

%\smallskip
\subsubsection{General cases}

In reality, the scale-free mass model is a rather artificial
abstraction.  Never the less, equation (\ref{eq:mm}) is actually valid
for any mass profile and arbitrary weights. Hence with
$h=r^{\alphap-1}$ and further assuming that the surface term vanishes
as $r_\mathrm{out}\rightarrow\infty$, we have that
\begin{equation}\label{eq:malp}
G\left\langle r^{\alphap-1}M\right\rangle=
(1+\alphap)\left\langle v_r^2r^\alphap\right\rangle
+\left\langle v_\mathrm t^2r^\alphap\right\rangle.
\end{equation}
We have used the symbol $\alphap$ to emphasize the fact that this needs
not be the same as the power-law index of the potential.  That is to
say, the scale-free estimator on the right-hand side actually returns
the radial moments of any given (not necessarily scale-free) mass
profile.  Although it is clear that the left-hand side is directly
related to the total mass, the specific constant factor associated
with it in general depends on both the mass profile and the tracer
density profile. There are some exceptions that are worthy of note
here. The first is the case of the power-law mass profile with $\alpha
= \alphap$, for which $\langle r^{\alphap-1}M(r)\rangle =\langle
r_\mathrm{out}^{\alpha-1}M_\mathrm{out}\rangle
=r_\mathrm{out}^{\alpha-1}M_\mathrm{out}$ is constant regardless of
the tracer density profile. The second is the self-consistent
gravitating system for which the $\alphap=1$ case results in
\begin{multline}
\left\langle M\right\rangle
=\frac{4\pi}{N_\mathrm{tot}}\!\int_0^\infty\!\dm r\,\nu r^2 M
=\frac1{M_\mathrm{tot}}\!\int_0^\infty\!\dm r\,\frac{\dm M}{\dm r}M
\\=\frac1{2M_\mathrm{tot}}\!\int_0^\infty\!\dm r\,\frac{\dm M^2}{\dm r}
=\frac{M_\mathrm{tot}}2
\end{multline}
because $\dm M/\dm r=4\pi r^2\rho$ where $\rho$ is the mass density
and the ratio $\rho/\nu=M_\mathrm{tot}/N_\mathrm{tot}$ is constant.
(By contrast, for a central $M_\bullet$ point-mass, we would have
$\left\langle M\right\rangle=\left\langle M_\bullet\right\rangle
=M_\bullet$.) An additional general result deduced from
equation (\ref{eq:malp}) is for $\alphap=0$, that is,
\begin{equation}\label{eq:vir}
\left\langle\frac{GM}r\right\rangle=
\left\langle v_r^2\right\rangle+\left\langle v_\mathrm t^2\right\rangle
=\langle v^2\rangle,
\end{equation}
which is simply %in fact nothing more than
the scalar virial theorem for a %non-rotating
spherical system.

\begin{table}
\begin{center}
\caption{Mean values $\langle v_r^2r^\alpha\rangle$ for
satellites in Table~1 of \citetalias{WEA}.}
\label{tab:los}
\begin{tabular}{lccc}\hline
$r_\mathrm{out}$& 300~kpc & 200~kpc & 100~kpc \\
&\multicolumn3c{($\times 10^{11}~\msun$)}\\\hline
$G^{-1}r_\mathrm{out}^{1-\alpha}\langle v_r^2r^\alpha\rangle$&
$2.86\pm0.87$ & $1.75\pm0.40$ & $1.20\pm0.22$
\\(w/o Leo I)& $2.09\pm0.42$ & &
\\(w/o Leo I, Her)& $1.76\pm0.27$ & $1.45\pm0.28$ &
\\\hline\end{tabular}
\end{center}
%\smallskip
Here $\alpha=0.55$ following \citetalias{WEA}.
Only satellites within $r\le r_\mathrm{out}$ are counted. These
consist of 14 satellites with $r_\mathrm{out}=100~\mbox{kpc}$,
21 (including Her) with $r_\mathrm{out}=200~\mbox{kpc}$,
and 24 (including Leo I and Her) with $r_\mathrm{out}=300~\mbox{kpc}$.
The quoted uncertainties, which range from $\sim15$
to $\sim30\%$, are purely due to the scatter, i.e., %and given by
$\sigma/N^{1/2}=s/\sqrt{N-1}$ where $\sigma^2$ and $s^2$ are
the unbiased and biased sample variance estimate and
$N$ is the number of data points.
\end{table}

%\bigskip
\section{The Mass of the Milky Way}
\subsection{Present Proper Motions of the Satellites}

\citetalias{WEA} have used combinations of the averages in the form of
$\langle v_r^2r^\alpha\rangle$ and $\langle
v^2r^\alpha\rangle=(3-2\beta)\langle v_r^2r^\alpha\rangle$ to estimate
the mass of the Galaxy from the kinematics of its satellites including
their proper motions. Whilst their approach of incorporating proper
motions has a valid justification in that the anisotropy value derived
by combining the line-of-sight (los) and proper motions appears to be
severely biased towards tangential motions, it is still instructive to
use the consistent mass estimator without a priori assumptions on the
anisotropy.

Here we use equation (\ref{eq:mpp}) to combine the los and
proper motion data gathered in \citetalias{WEA} and estimate the
Galactic mass. We first note the scaling relation for
observable combinations of kinematic properties
($\mbox{\muasa}\equiv$ micro-arcseconds per year):
%
%\begin{subequations}
\begin{equation}\begin{split}%\begin{align}
\frac{v_r^2r^\alpha}{Gr_\mathrm{out}^{\alpha-1}}&=
2.325\cdot10^9~\msun
\biggl(\dfrac{r_\mathrm{out}}{100~\mbox{kpc}}\biggr)^{1-\alpha}
\biggl(\dfrac r{100~\mbox{kpc}}\biggr)^\alpha
\biggl(\dfrac{v_\ell}{10~\mbox{km s$^{-1}$}}\biggr)^2;
\\\frac{v_\mathrm t^2r^\alpha}{Gr_\mathrm{out}^{\alpha-1}}&=
5.225\cdot10^8~\msun
\biggl(\dfrac{r_\mathrm{out}}{100~\mbox{kpc}}\biggr)^{1-\alpha}
\biggl(\dfrac r{100~\mbox{kpc}}\biggr)^{\alpha+2}
\biggl(\dfrac\mu{10~\mbox{\muasa}}\biggr)^2.
\end{split}\end{equation}%\end{align}
%\end{subequations}
%
Here, $v_\ell$ and $\mu$ are the los velocity and
the proper motion of the satellite, respectively. For the satellites
considered here, the Galactocentic distances are large enough
to ignore the sun being off-centre in the Galaxy.
We shall therefore assume that $v_r=v_\ell$ and $\bmath v_\mathrm t=r\bmu$,
and the use of equation (\ref{eq:mpp}) is appropriate.

Table~\ref{tab:los} lists the contribution to %the mass estimator in
equation (\ref{eq:mpp}) of the radial motions of satellites assembled
in Table~1 of \citetalias{WEA} with $\alpha=0.55$, which is
the `best-fitting' value of the power-law index to the potential
of a typical NFW halo as determined by \citetalias{WEA}.
The uncertainties quoted are purely
statistical and typically of about $\sim20$ per cent, comparable to
$\sim25$ per cent derived via Monte Carlo simulations in
\citetalias{WEA}. Assuming isotropy and adopting the tracer power-law slope
$\gamma=2.6$, the Galactic mass based on these data can be estimated
by multiplying these by a factor of 3.15, the results of which
basically reproduce those reported in Table~3 of \citetalias{WEA}.

\begin{table}
\begin{center}
\caption{Numerical contribution to the proper motion
mass estimator $G^{-1}r_\mathrm{out}^{1-\alpha}v_\mathrm t^2r^\alpha$
%=G^{-1}r_\mathrm{out}^{1-\alpha}\mu^2r^{2+\alpha}$
for satellites in Table~4 of \citetalias{WEA}.}
\label{tab:pm}
\begin{tabular}{lcc}\hline
$r_\mathrm{out}$& 200~kpc & 100~kpc \\
&\multicolumn2c{($\times 10^{11}~\msun$)}\\\hline
Car& $\p5.3\pm\p3.6$ &
\\Dra& $90.6\pm47.1$ & $66.3\pm34.5$
\\For& $60.6\pm\p9.4$ &
\\MCs& $48.5\pm\p2.4$ & $35.5\pm\p1.8$
\\Scl& $0.43\pm1.20$ & $0.31\pm0.88$
\\Sex& $\p4.1\pm12.2$ & $\p3.0\pm\p9.0$
\\UMi& $10.9\pm\p6.7$ & $\p8.0\pm\p4.9$
\\\hline\end{tabular}
\end{center}
%\smallskip
Here again $\alpha=0.55$.
All listed satellites have $r<100~\mbox{kpc}$ except for
Carina ($r=102~\mbox{kpc}$) and Fornax ($r=140~\mbox{kpc}$). For the
Magellanic Clouds, we choose $r=50~\mbox{kpc}$ for the barycentre
(assuming the mass ratio of $\text{LMC/SMC}\sim11$). The quoted errors
are the directly propagated uncertainties of the proper motion measurements
assuming no correlation between the two components of the proper motions.
\end{table}

Table~\ref{tab:pm} on the other hand provides us with the individual
contributions to the mass estimator of the proper motions measured for
seven Galactic satellites reported in Table~4 of \citetalias{WEA}. In
contrast to the los velocities, the uncertainties associated
with each measurement are rather large.  This not only contributes
additional uncertainties in the final mass estimate, but also
complicates the proper weighting scheme appropriate for the average
related to the mass estimator.  In Table~\ref{tab:mass}, we present
the result calculated with equation (\ref{eq:mpp}). For the average of
the proper motion contributions, the results based on both the
unweighted arithmetic mean and the mean weighted by the reciprocal of
the observational variance (i.e., $\sigma^{-2}$ where $\sigma$ is the
associated uncertainty reported in Table~\ref{tab:pm}) are
given. Compared to the results based on the los motions
alone, the final statistical uncertainties are quite larger ranging
from $\sim50$ per cent to a factor of two.  More seriously, the
results vary in significant amounts depending on the weighting scheme
and the choice of outliers (which may be due to observational errors
and/or true unbound objects), although they are formally consistent
with one another given large error bars.  It appears that the safest
conclusion that can be drawn at this point is that the existing proper
motion data do not provide any meaningful discriminating information
on the mass of the Milky Way without any a priori assumption on the
motions of the Galactic satellites as a whole or the particular
target.

\begin{table}
\begin{center}
\caption{Mass of the Milky Way}
\label{tab:mass}
\begin{tabular}{lccc}\hline
$r_\mathrm{out}$& \multicolumn2c{200~kpc} & 100~kpc \\
&&(w/o Her)&\\\hline
unweighted& $33.5\pm15.4$ & $33.2\pm15.4$ & $24.0\pm14.9$ \\
(w/o Dra)& $23.7\pm11.1$ & $23.3\pm11.1$ & $13.1\pm\p8.6$ \\
(w/o MCs)& $30.7\pm17.9$ & $30.4\pm17.9$ & $20.8\pm18.8$ \\
(w/o Dra,MCs)& $18.3\pm11.9$ & $18.0\pm11.9$ & $\p5.2\pm\p4.8$ \\\hline
$\sigma^{-2}$-weighted& $12.1\pm10.6$ & $11.8\pm10.6$ & $\p8.6\pm10.4$ \\
(w/o Dra)& $12.1\pm11.4$ & $11.8\pm11.4$ & $\p8.6\pm11.6$ \\
(w/o MCs)& $\p4.1\pm\p6.5$ & $\p3.8\pm\p6.5$ & $\p2.0\pm\p4.9$ \\
(w/o Dra,MCs)& $\p4.1\pm\p6.8$ & $\p3.7\pm\p6.8$ & $\p2.0\pm\p4.2$
\\\hline\end{tabular}
\end{center}
%\smallskip
The estimator uses equation (\ref{eq:mpp}) with
$\alpha=0.55$ and $\gamma=2.6$. The mass is in the units of
$10^{11}~\msun$. The uncertainties are purely statistical
incorporating both scatters and the measurement uncertainties found
in Table~\ref{tab:pm}.
\end{table}

%\medskip
\subsection{The Age of \emph{Gaia}}

Let us now estimate the benefits that {\it Gaia} will bring. We
do this by constructing mock datasets of los velocities and
proper motions for sets of dwarf galaxy satellites extracted from a
suite of simulations, the {\it Galaxies-Intergalactic Medium
Interaction Calculation}, %({\it GIMIC}) simulations
which are described in detail in \citet{Ca09}.
They consist of a set of hydrodynamical resimulations of
five nearly spherical regions ($\sim 20 h^{-1}\ \mbox{Mpc}$ in radius) extracted from the {\it Millennium Simulation}
\citep{Sp05}. \citet{De11} extracted a set of galaxies that resemble
the Milky Way. The catalogue consists of 431 parent haloes and 4864
associated satellite galaxies.

\begin{figure*}
\begin{center}
\includegraphics[width=0.49\textwidth]{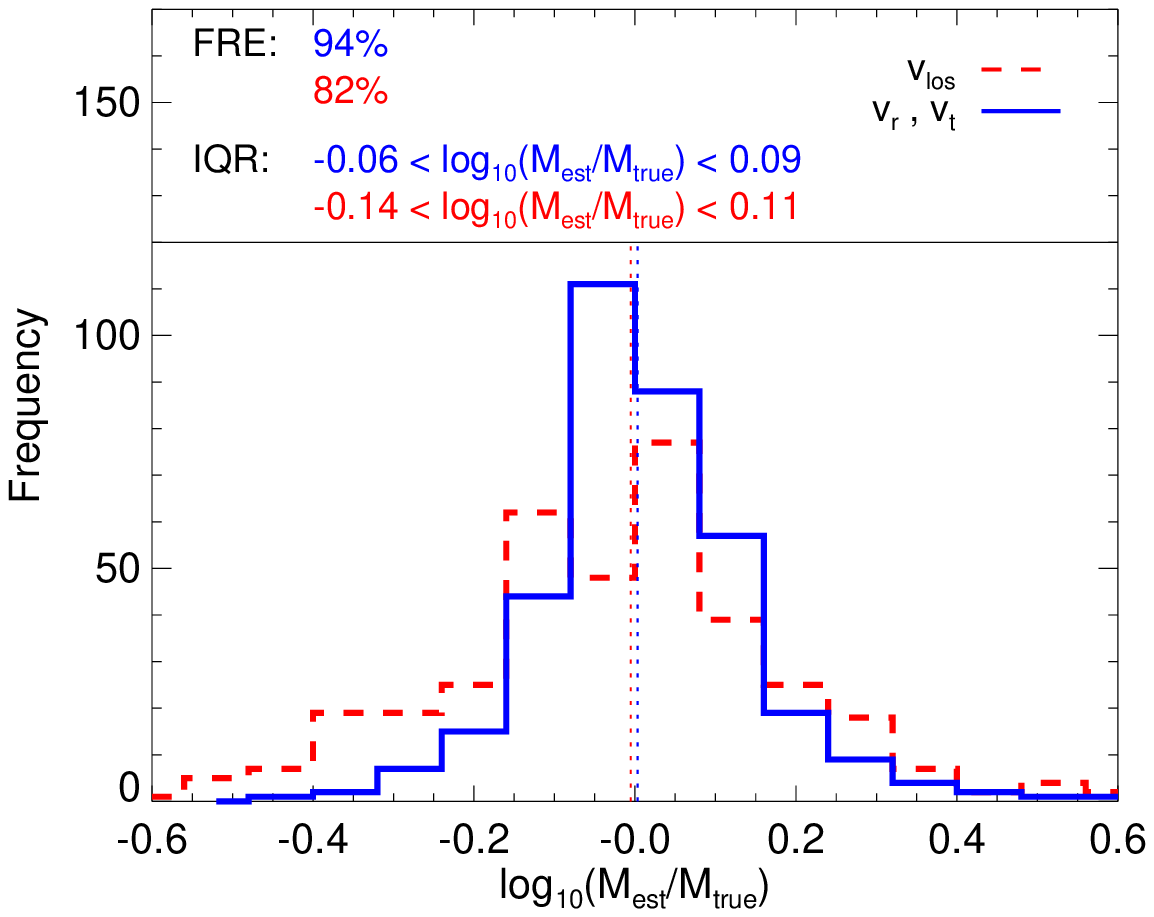}
\includegraphics[width=0.49\textwidth]{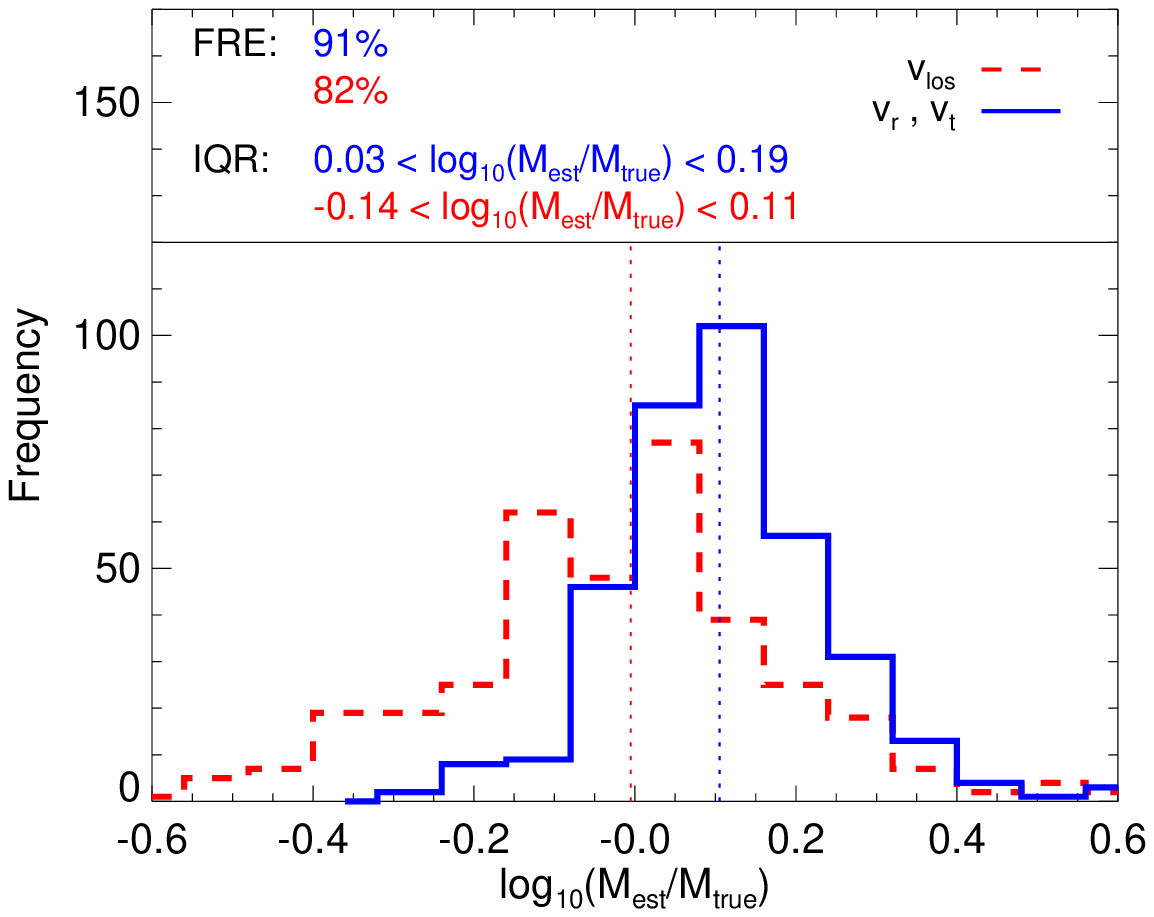}
\end{center}
\caption{Left: Histograms showing the ratio of the estimated mass to the
  true mass for the two different mass estimators. The dashed histogram
  uses sets of radial velocities and distances with the
  tracer mass estimator of \citetalias{WEA}. The solid histogram uses sets
  full space velocities and distances, with mass estimator of
  eq.~(\ref{eq:mpp}). The FRE and IQR are the fraction
  of reasonable estimates and the inter-quartile range as
  defined in the text. The means of the distributions are shown by
  vertical dotted lines. Right: As the left panel, but now the solid
  histogram uses eq.~(\ref{eq:mep}) neglecting the surface term.}
\label{fig:sims}
\end{figure*}

The mass estimators provide the total mass within the radius of the
farthest tracer ($\approx r_\mathrm{out}$). We compute the `true' mass
within the virial radius of each halo and compare to the masses found
via our estimator. We use all satellites,
but check that our results are not significantly affected when only
luminous satellites are included. Figure~\ref{fig:sims} refers to data
sets of radial velocities and true distances (dashed histogram), and
radial velocities, proper motions and projected distances (solid
histogram). An uncertainty of 100~\muasa for each star is
assumed; for comparison, {\it Gaia}'s targeted accuracies are 10~\muasa
at $G\sim15$ and 150~\muasa at $G\sim20$ \citep{Li08}.
For each tangential velocity measurement, a
Gaussian distribution centred on the `real measurement' with error is
derived from the proper motion (and so depends on distance).  For each
satellite, it is assumed that {\it Gaia} will detect 100 stars, which is
reasonable enough for the case of the satellites of the Milky Way
\citep{WE99}. The left panel of Figure~\ref{fig:sims} uses the estimator
given by equation (\ref{eq:mpp}), the right panel uses equation
(\ref{eq:mep}) neglecting the surface term.

We quantify the improvement by means of two statistical measures
as originally introduced by \citetalias{Ev11}.
First, we define the Fraction of Reasonable Estimates (FRE)
as the fraction of estimates within the factor of two of the true mass
\citep[see also][]{De11}.
We also give the Inter Quartile Range (IQR)
of the mass estimates, which gives a good indication of the spread.
We see that at {\it Gaia}'s accuracy, the proper motions can contribute
significantly to an improved estimate of the mass of the Milky Way.
Even with the small number of satellites, the inclusion of proper
motion data via the estimator of equation (\ref{eq:mpp}) means that
93 per cent of estimates are within a factor of two, as opposed to
82 per cent with radial velocities alone. The solid histogram is also
more narrow than the dashed, as indicated by a smaller IQR
of $-0.05<\log(M_\mathrm{est}/M_\mathrm{true})<0.09$. Notice from the right
panel that the estimator of equation (\ref{eq:mep}) also performs very
well with regard to the FRE, but the neglect
of the boundary term causes an offset so the estimator is biased.

As is noted by \citetalias{Ev11}, there are a number of ways in which
mock catalogues from simulations differ from the assumptions used to
derive the estimators. For example, dark haloes are not generally
spherical, infall continues to the present day, and the observed
satellites are not necessarily virialized and well-described by an
equilibrium distribution.
Therefore the performance of the mass estimators is very encouraging.

%\vfill
%\bigskip
\section{Projected Data with Proper Motions}

In most practical cases, the kinematic data available are limited by
projection. Hence, the naturally observable velocity dispersions are
those along the line of sight and of proper motions. In addition, for
most dynamical systems of interest, their spatial extents are
typically much smaller than their distances from us as a whole.  On
the positive side, this indicates that, from the observer's
perspective, the observed los velocities $v_\ell$ basically
correspond to the coordinate velocity component $v_z$ if we have
chosen the cylindrical polar coordinate centred at the centre of the
tracer population and the positive $z$-axis running from the centre to
the observer. In addition, the remaining two orthogonal components of
the velocity $\bmath v_\wp=(v_R,v_\phi)$ are linearly proportional to
the two components of the proper motion with $\bmath v_\wp=s\bmu$
where $s$ is the distance to the system as a whole and $\bmu$ is the
proper motion vector.

On the negative side however, the radial distance of any individual
tracer to the centre of the system is very difficult to determine
precisely. In particular, since the absolute uncertainty ($\delta r$)
of the radial distance ($r$) measurement is in the same order as
the uncertainty ($\delta s$) of the individyal distance ($s$) measurement,
i.e., $\delta r\sim\delta s$ (NB., the error in the distance measurement
typically dominates that in the angular separation whilst the precise
determination of the distance to the centre only improves the precision
by $\sim\sqrt2$), the fractional uncertainty of the radial distance
is roughly estimated to be $(\delta r/r)\sim(\delta s/s)(s/r)
\sim(\delta s/s)/\varphi$ where $\varphi\sim r/s$ is of order the
angular extent of the system. In other words, $\sim10$ per cent
estimates of the radial distances for tracers in a degree-sized system
require the individual distance measurement in a better than $\sim0.2$
per cent precision. Consequently, the distance handle associated with
the tracer is likely to be limited to the projected (linear
or angular, depending on whether the distance to the system as a whole
is known or not) separation.  In the following, we develop the
projected mass estimators similar to those of \citet{BT81} and
\citet{He85} using $v_\ell$ and $R$ as observables but further
augmented by including the transverse velocity $\bmath v_\wp$ derived
from the proper motion.  We note that this problem has been partly
addressed by \citet{LM89}.

%\medskip
\subsection{The Projected Mass Estimator with Transverse Velocities}

Let us first consider the coordinate system with
its origin located at the centre of the system and the positive
$z$-axis running from the origin to us. Then, the tracer velocity
components in the spherical polar coordinate $(v_r,v_\theta,v_\phi)$
transform to those in the cylindrical coordinate $(v_R,v_\phi,v_z)$
according to
\begin{equation}
v_R=v_r\sin\theta+v_\theta\cos\theta\,;\qquad
v_z=v_r\cos\theta-v_\theta\sin\theta.
\end{equation}
If the tracers are described by a generic two-integral distribution
that is non-rotating and spherically symmetric, then the
velocity dispersion tensor is diagonalized in the spherical polar
coordinate with $\sigma_{\theta\theta}=\sigma_{\phi\phi}$.
These transform to
the velocity dispersion tensor in the cylindrical polar coordinate via
\begin{equation}\label{eq:sigt}\begin{split}
\sigma_{RR}^2&=\sigma_r^2\sin^2\!\theta+\sigma_\theta^2\cos^2\!\theta
=(1-\beta\cos^2\!\theta)\,\sigma_r^2
\\\sigma_{\phi\phi}^2&=\sigma_\theta^2=(1-\beta)\,\sigma_r^2
\\\sigma_{zz}^2&=\sigma_r^2\cos^2\!\theta+\sigma_\theta^2\sin^2\!\theta
=(1-\beta\sin^2\!\theta)\,\sigma_r^2
\\\sigma_{Rz}^2&=(\sigma_r^2-\sigma_\theta^2)\,\sin\theta\cos\theta
=\beta\sin\theta\cos\theta\,\sigma_r^2,
\end{split}\end{equation}
where $\sigma_r^2=\sigma_{rr}^2$ and
$\sigma_\theta^2=\sigma_{\theta\theta}^2$, and $\beta$ is given by
equation (\ref{eq:beta}).

Using equation (\ref{eq:sigt}), it is trivial to find
for $\alpha>-2$ that
\begin{subequations}\label{eq:vRa}\begin{align}\begin{split}
\left\langle v_R^2R^\alpha\right\rangle
&=\frac{2\pi}{N_\mathrm{tot}}\!\int_0^\infty\!\dm r\,r^2\nu(r)
\!\int_0^\pi\!\dm\theta\,\sin\theta\,
\sigma_{RR}^2r^\alpha\sin^\alpha\!\theta
\\&=\frac{2\pi}{N_\mathrm{tot}}\!\int_0^\infty\!\dm r\,\nu\,r^{2+\alpha}
\left(\sigma_r^2\mathrm B_{2+\frac\alpha2,\frac12}
+\sigma_\theta^2\mathrm B_{1+\frac\alpha2,\frac32}\right)
\\&=B_\alpha\left[(2+\alpha)\langle v_r^2r^\alpha\rangle
+\langle v_\theta^2r^\alpha\rangle\right]
\end{split}\\
\left\langle v_\phi^2R^\alpha\right\rangle
&=(3+\alpha)B_\alpha\langle v_\theta^2r^\alpha\rangle
\\
\left\langle v_z^2R^\alpha\right\rangle
&=B_\alpha\left[\langle v_r^2r^\alpha\rangle
+(2+\alpha)\langle v_\theta^2r^\alpha\rangle\right]
\end{align}
where $\mathrm B_{a,b}=\Gamma(a)\Gamma(b)/\Gamma(a+b)$ and $\Gamma(x)$
are the beta and gamma function and
\begin{equation}
B_\alpha\equiv\tfrac12\mathrm B_{1+\frac\alpha2,\frac32}=
\frac{\sqrt\pi\Gamma(1+\frac\alpha2)}{4\Gamma\bigl(\frac{5+\alpha}2\bigr)}.
\end{equation}\end{subequations}
The result extends to the $\alpha=-2$ case
(homogeneous sphere) with
\begin{subequations}\label{eq:alp2}
\begin{equation}\begin{split}
\left\langle v_X^2\deltaup(R^2)\right\rangle
&=\frac{2\pi}{N_\mathrm{tot}}\!
\int_{-\infty}^\infty\!\dm z\!\int_0^\infty\!\dm R\,R
\nu\sigma_{XX}^2\deltaup(R^2)
\\&=\frac\pi{N_\mathrm{tot}}\!\int_{-\infty}^\infty\!\dm z\,
\nu\sigma_{XX}^2\bigr\rvert_{R=0,z}
\end{split}\end{equation}
where $\deltaup(x)$ is the Dirac delta. Since $(R=0,z>0)$
and $(R=0,z<0)$ correspond to $(r=z,\theta=0)$ and
$(r=\lvert z\rvert,\theta=\pi)$,
we find from equation (\ref{eq:sigt}) that
\begin{equation}\begin{split}
\left\langle v_R^2\deltaup(R^2)\right\rangle
&=\frac{2\pi}{N_\mathrm{tot}}\!
\int_0^\infty\!\dm r\,\nu\sigma_\theta^2
=\tfrac12{\langle v_\theta^2r^{-2}\rangle}\\
\left\langle v_\phi^2\deltaup(R^2)\right\rangle
&=\frac{2\pi}{N_\mathrm{tot}}\!
\int_0^\infty\!\dm r\,\nu\sigma_\theta^2
=\tfrac12{\langle v_\theta^2r^{-2}\rangle}\\
\left\langle v_z^2\deltaup(R^2)\right\rangle
&=\frac{2\pi}{N_\mathrm{tot}}\!
\int_0^\infty\!\dm r\,\nu\sigma_r^2
=\tfrac12{\langle v_r^2r^{-2}\rangle}.
\end{split}\end{equation}
\end{subequations}
This result is identified with the formal limit of
equation (\ref{eq:vRa}) as $\alpha\rightarrow{-2}^+$ once
we note that
$\lim_{\alpha\rightarrow{-2}^+}R^\alpha/\Gamma(1+\frac\alpha2)
=\deltaup(R^2)$.

For an observer located sufficiently far away,
the axial and sectional velocities are basically the los and
the transverse velocities, respectively.
%That is, $\bmath v_\wp=s\bmu=(v_R,v_\phi)$,
%$v_\wp=|\bmath v_\wp|=s\mu=\sqrt{v_R^2+v_\phi^2}$ and $v_\ell=v_z$.
It follows that $\left\langle v_\ell^2R^\alpha\right\rangle=
\left\langle v_z^2R^\alpha\right\rangle$ and
\begin{equation}
\left\langle v_\wp^2R^\alpha\right\rangle
=B_\alpha
\left[(2+\alpha)\langle v_r^2r^\alpha\rangle
+(4+\alpha)\langle v_\theta^2r^\alpha\rangle\right].
\end{equation}
If the two components of the transverse velocities are
separately measurable, we can derive the right-hand of
equation (\ref{eq:malp}) from
\begin{subequations}
\begin{equation}
(1+\alpha)\left\langle v_R^2R^\alpha\right\rangle
+\left\langle v_\phi^2R^\alpha\right\rangle
=(2+\alpha)B_\alpha
\left[(1+\alpha)\langle v_r^2r^\alpha\rangle
+2\langle v_\theta^2r^\alpha\rangle\right].
\end{equation}
On the other hand, the same may be found by combining the los and
the two-dimensional transverse velocity dispersions
\begin{equation}
\left\langle v_\wp^2R^\alpha\right\rangle
-\left\langle v_\ell^2R^\alpha\right\rangle
=B_\alpha
\left[(1+\alpha)\langle v_r^2r^\alpha\rangle
+2\langle v_\theta^2r^\alpha\rangle\right].
\end{equation}\end{subequations}
Directly substituting these in equation (\ref{eq:malp}) then leads to
\begin{subequations}
\begin{equation}\label{eq:pmep}
GB_\alpha\left\langle r^{\alpha-1}M\right\rangle
=\frac{(1+\alpha)\left\langle v_R^2R^\alpha\right\rangle
+\left\langle v_\phi^2R^\alpha\right\rangle}{2+\alpha}
=\left\langle v_\wp^2R^\alpha\right\rangle
-\left\langle v_\ell^2R^\alpha\right\rangle,
\end{equation}
which is again valid for $\alpha>-2$.
We note that equation (\ref{eq:pmep}) with $\alpha=-1,0,1,2$
are already found in \citet{LM89}. If $\alpha=-2$, then strictly
the average is not convergent, but using equation (\ref{eq:alp2}), we
find the proper limiting equation, namely,
\begin{equation}
\tfrac12G\left\langle r^{-3}M\right\rangle
=\left\langle v_\wp^2\deltaup(R^2)\right\rangle
-\left\langle v_\ell^2\deltaup(R^2)\right\rangle,
\end{equation}
whereas
$\left\langle v_R^2\deltaup(R^2)\right\rangle=
\left\langle v_\phi^2\deltaup(R^2)\right\rangle$.
\end{subequations}
Provided that the distance $s$ to the system as a whole
is known independently, equation (\ref{eq:pmep}) is
the proper extension of the projected mass estimator
to include the information from proper motions.
Note that, if the vectorial dispersion of the proper motions is known,
incorporating the los velocity dispersions
in fact over-constrains the system under the assumption of
the two-integral spherically-symmetric distribution.

\begin{figure*}
\begin{center}
\includegraphics[width=0.49\textwidth]{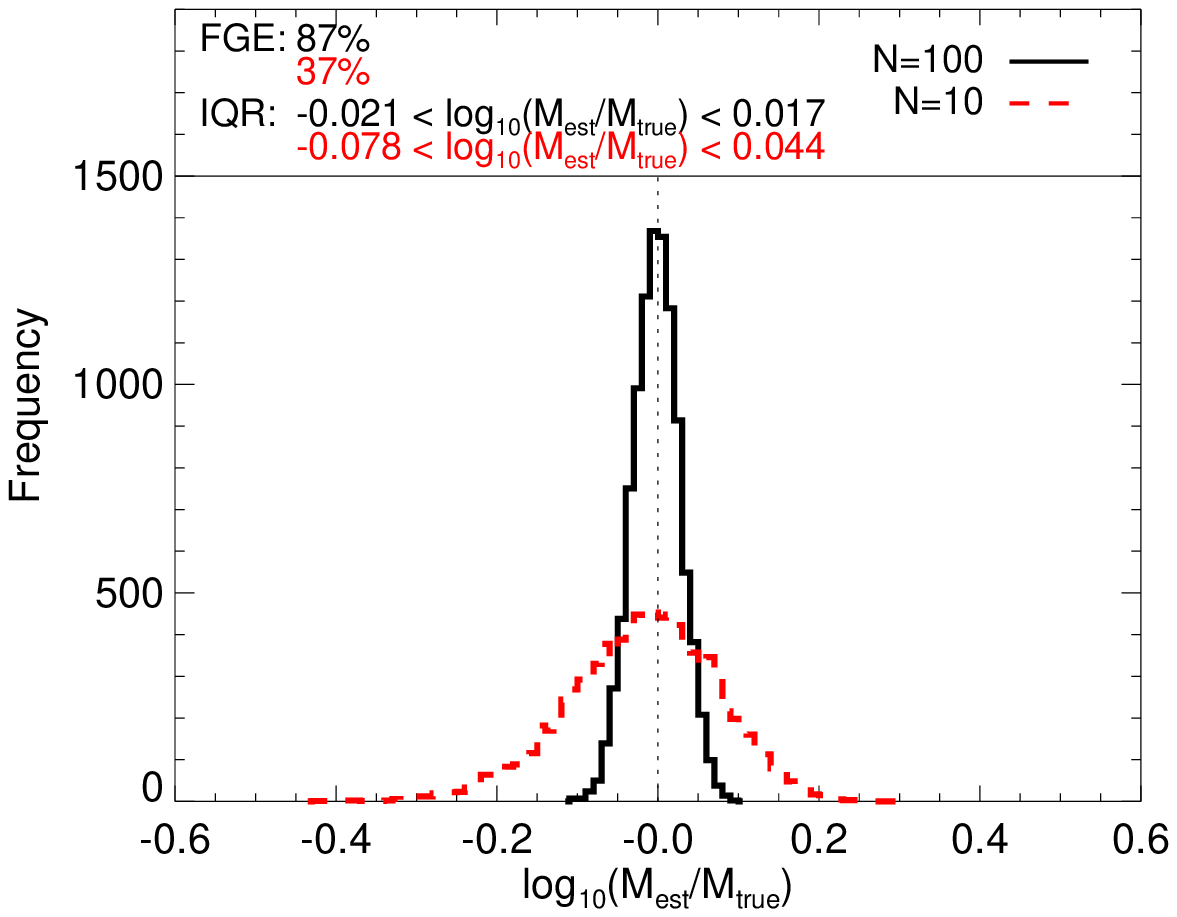}
\includegraphics[width=0.49\textwidth]{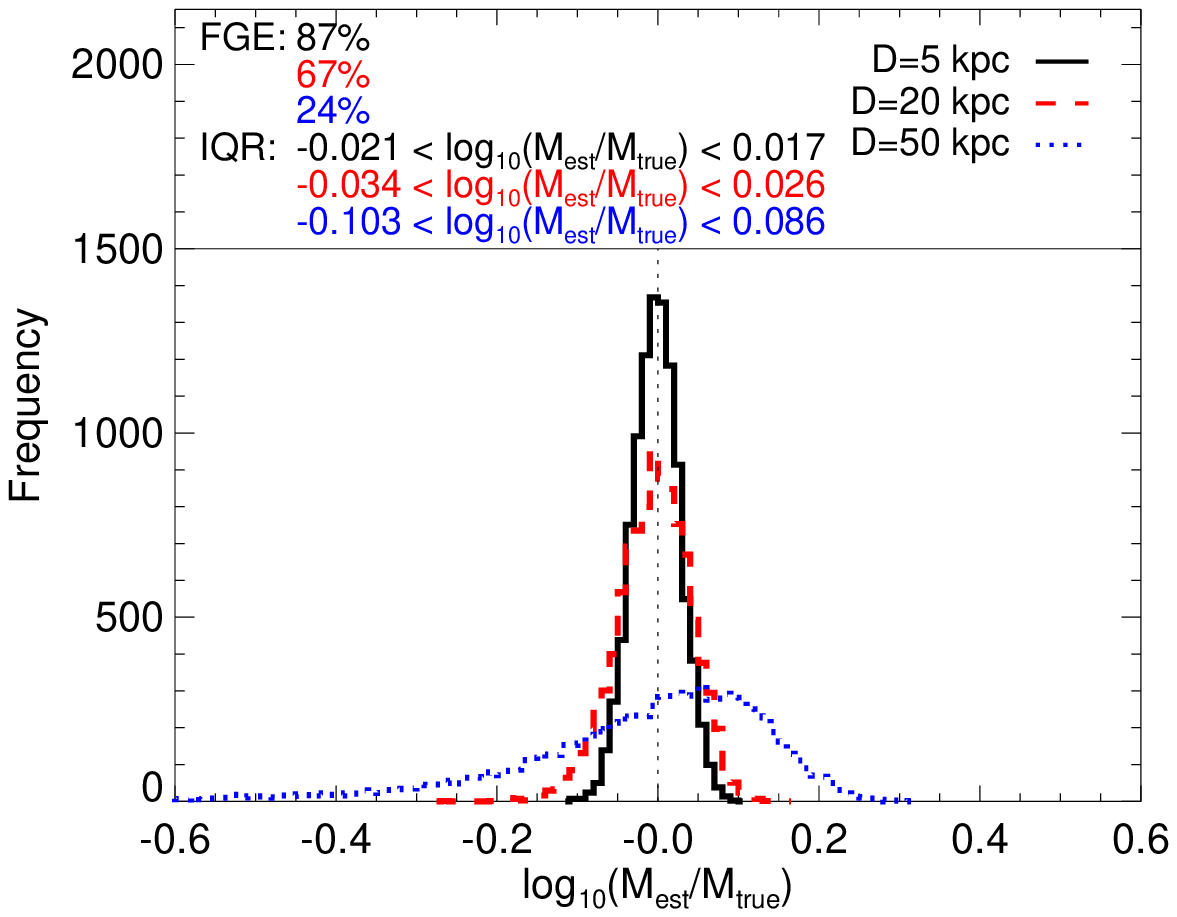}
\end{center}
\caption{Left: Estimates of enclosed mass of a globular cluster using
  10 (dashed) and 100 (solid) tracers from $10^5$ Monte Carlo
  realizations. Proper motion errors of 100~\muasa and
  a distance of 5 kpc are assumed.  Right: Estimates of enclosed mass
  as a function of globular cluster distance, namely 5 (solid), 20
  (dashed) or 50 (dotted) kpc, assuming 100 stars can be measured per
  globular cluster. Model has parameters $\alpha=0$ and $\gamma=3$. }
\label{fig:simstwo}
\end{figure*}

%\smallskip
\subsubsection{Mass measurements of globular clusters}

We can test these formulae in the context of applications.
A possible class of targets would be nearby dwarf spheroidal/elliptical
galaxies, similar to the examples found in \citetalias{Ev11}. However,
the proper motion dispersion for an imaginary satellite of mass
$10^7\ \msun$ and radius 0.5~kpc located at 50~kpc
is about $\sim40~\mbox{\muasa}$. Even with {\it Gaia},
these will be challenging observations even if there exist
large enough samples of sufficiently luminous tracers,
which is itself unclear in any case.
Alternatively, we can consider mass estimates of Galactic globular
clusters. In fact, for a few bright clusters, there already exist
very good internal proper motion data sets \citep[e.g.,][]{AvdM10}.
Yet the existing data for these are sufficiently large and of high
quality so that the complete dynamical modelling is technically
feasible \citep[see e.g.,][]{vdMA10} -- note that there is no degeneracy
involving anisotropy with the proper motion data incorporated.
In principle, we may still utilize these for the projected mass estimator
but its application is clearly didactic at best. Instead we
investigate {\it Gaia}'s capability regarding
possible applications of projected mass estimators
for `typical' globular clusters in varying distances.

In contrast to dwarf satellites, the main challenge of globular
cluster observations is expected to be crowding, and so
the choice of the tracer subsample is important for any practical
application of mass estimators.
Blue horizontal branch (BHB) stars have a $G$-band magnitude
of about $\sim19$ at 50~kpc.  So, {\it Gaia} will be able to
detect a BHB star out to 50~kpc and measure their proper motions with
accuracies of better than 100~\muasa. By contrast,
radial velocities will usually be missing, as {\it Gaia}'s radial
velocity spectrograph has a limiting magnitude of $G \approx 17$.
Most globular clusters harbour significant populations of BHB
stars \citep[e.g.,][]{Br05}, and so equation (\ref{eq:pmep}) can be
exploited to measure the mass of globular clusters in the Milky Way
from proper motions provided by {\it Gaia}. To test the feasibility,
we use the the self-consistent distributions of power-law densities
with $\rho \propto r^{-3}$ in logarithmic potentials \citep{Wh81,Ev97}
to generate mock datasets of proper motions and projected distances,
degraded with the {\it Gaia} errors. The estimated mass computed via
equation (\ref{eq:pmep}) can then be compared to the known true
mass. Figure~\ref{fig:simstwo} shows the results of Monte Carlo
simulations.  In the left panel, the globular cluster is assumed to
lie at a distance of 5~kpc and the sample size is 10 or 100 tracers
with projected positions and proper motions. As the histograms are
quite tight, we have replaced FRE with
the Fraction of Good Estimates (FGE), where ``Good'' means that the mass
estimate is within 10 per cent of the true mass. We see that 10
tracers are too small for reliable results, but 100 tracers are ample.
The right panel shows the estimated masses as a function of distance,
with the globular cluster assumed to lie at 5, 20 and 50~kpc
respectively. In each case, 100 tracers are used, and we see that the
results are still acceptable at distances of 20~kpc, with 67 per cent
of the mass estimates being good. With increasing distance, however,
the accuracy degrades so that by 50~kpc, only 24 per cent of the
estimates are good.

%\medskip
\subsection{The Projected Mass Estimator with Proper Motions}

Now we consider the case that the distance $s$ is also unknown.
It can in principle be measured by matching
the los velocity and the proper motion dispersions.
Specifically, the last equality of equation (\ref{eq:pmep}) is
rearranged to
\begin{equation}
(2+\alpha)\left\langle v_\ell^2R^\alpha\right\rangle
=\left\langle v_R^2R^\alpha\right\rangle
+(1+\alpha)\left\langle v_\phi^2R^\alpha\right\rangle.
\end{equation}
With $R=s\varphi$ and $\bmath v_\wp=s\bmu$ where
$\varphi$ is the angular separation from the centre,
the distance $s$ is found from direct observables as in
\begin{equation}\label{eq:s2}
s^2=\frac{(2+\alpha)\left\langle v_\ell^2\varphi^\alpha\right\rangle}
{\left\langle\mu_R^2\varphi^\alpha\right\rangle
+(1+\alpha)\left\langle\mu_\phi^2\varphi^\alpha\right\rangle},
\end{equation}
where $\mu_R$ and $\mu_\phi$ are
the radial and azimuthal projections of the proper motion.
Note that the $\alpha=0$ case reduces to
$s^2\langle\mu^2\rangle=\langle v_\ell^2\rangle$ irrespective of
the anisotropy, which corresponds to the classical
statistical parallax technique of measuring distance. This is because
only in the $\alpha =0$ case are the velocity distributions
independent of position, which usually assumed in statistical parallax
applications \citep[see e.g.,][]{SA}.

At the same time, we can also solve for the anisotropy parameter
\begin{equation}\label{eq:beff}
\beta_\mathrm{eff}=1-
\frac{\langle v_\theta^2r^\alpha\rangle}{\langle v_r^2r^\alpha\rangle}
=\frac{(3+\alpha)\left[\left\langle\mu_R^2\varphi^\alpha\right\rangle
-\left\langle\mu_\phi^2\varphi^\alpha\right\rangle\right]}
{(3+\alpha)\left\langle\mu_R^2\varphi^\alpha\right\rangle
-\left\langle\mu_\phi^2\varphi^\alpha\right\rangle}.
\end{equation}
The unique mass estimator that follows this is
\begin{equation}\label{eq:mproj}
GB_\alpha\left\langle r^{\alpha-1}M\right\rangle
=\frac{(1+\alpha)\left\langle\mu_R^2\varphi^\alpha\right\rangle
+\left\langle\mu_\phi^2\varphi^\alpha\right\rangle}
{\left\langle\mu_R^2\varphi^\alpha\right\rangle
+(1+\alpha)\left\langle\mu_\phi^2\varphi^\alpha\right\rangle}
\left\langle v_\ell^2R^\alpha\right\rangle
\end{equation}
with $\langle v_\ell^2R^\alpha\rangle
=s^\alpha\langle v_\ell^2\varphi^\alpha\rangle$.
This again is in fact equivalent to
equations (24) and (25) of \citetalias{AE11} with $\beta$ replaced by
equation (\ref{eq:beff}) and the boundary term dropped. If one preferred
the explicitly solved boundary term used in \citetalias{WEA}, one might
use their equations (26) and (27) with $\beta$ replaced by
the same equation (\ref{eq:beff}).
However, we note that estimators in
equations (\ref{eq:s2})-(\ref{eq:mproj}) are likely to be biased
if the sample means are used to estimate
for the averages in the right-hand sides
as these formulae involve a division by a mean whose corresponding
dispersion is finite non-zero.

%\bigskip
\section{Conclusions}

Hitherto, proper motion data have not normally been used in
measurements of the mass of nearby galaxies like the Milky Way and
M31. Proper motions are only available for a handful of satellites --
namely, Carina, Draco, Fornax, Sculptor, Sextans, Ursa Minor and the
Large and Small Magellanic Clouds for the Milky Way, together with
IC10 and M33 for M31. As these are difficult measurements, the errors
are often quite large. Additionally, agreement between different
investigators regarding the proper motions for Ursa Minor
\citep{Sc94,Pi05} and Sculptor \citep{Pi06,Wa08} is poor. Therefore,
there has been little attention given to their exploitation.

This situation is about to change. The {\it Gaia} satellite, to be
launched in 2013, will measure the positions and proper motions of
$\sim 10^9$ stars in our Galaxy and other members of the Local
Group. The proper motions have a targeted accuracy of between
10~\muasa (at $G\sim15$) and 150~\muasa (at $G\sim20$).
This therefore motivates detailed study of the use
of proper motions in mass estimators. Particularly in the case of the
Milky Way, the proper motions are crucial as the line-of-sight
velocity closely coincides with the radial velocity as measured from
the Galactic Centre. Hence, proper motions provide new information on
the velocity anisotropy that is not available in the line-of-sight
velocities alone.

Here we have shown how to solve the classical problem of estimating
the enclosed mass from the positions and space velocities of a set of
tracers. The main result is equation (\ref{eq:main}), which shows how
to compute the enclosed mass in terms of averages over the observable
velocity components for a given halo mass model. If, as is reasonable,
we use a scale-free assumption for the dark halo, then the mass
estimator reduces to equation (\ref{eq:mpp}).
%which can be recast as:
%
%\begin{equation}
%M_\mathrm{out} \approx  {(\alpha + \gamma -2)\langle v_r^2r^\alpha\rangle
%+\langle v_\mathrm t^2r^\alpha\rangle \over G r_\mathrm{out}^{\alpha-1}},
%\end{equation}
%
%where $\gamma$ is the power-law index of the (three-dimensional)
%tracer density profile.
Note values of $\alpha \sim 0.5$ for the power-law index for the potential
are good approximations to the NFW halo \citep{De11,Ev11}.

We have used these new formulae to demonstrate that the existing
proper motion data for the Milky Way satellites provide no
meaningful discriminating information on the mass of the Milky Way.
This though will change in the era of {\it Gaia}. Using mock
catalogues of satellites extracted from the simulations, we have shown
that our new mass estimator, when equipped with proper motions with
the targeted {\it Gaia} accuracies, significantly outperforms the
estimator used with line-of-sight velocities only. We have also given
the extension of the formulae to the instances where projected
positions and proper motions are available. These formulae provide a
feasible method for the {\it Gaia} satellite to measure the internal
masses of Milky Way globular clusters lying within 20~kpc.

As is usual for a spherical system, we have assumed no rotation
in the tracer population for our derivation of the mass estimators.
(Clearly, any rotating population cannot be spherically symmetric
although it may be spherical in density.) However, the tracer
velocities in our formulae appear as their second moments rather
than the dispersions/variances (in the rest frame of the potential).
That is, once the bulk \emph{linear} motion of the tracers is subtracted,
the estimators actually do not distinguish between the random
and coherent motions of the tracers. Hence, we argue that the formulae
as they are are still valid even if the population is in fact supported
through the mixture of the pressure and the rotation.

%\vfill
\section*{Acknowledgments}%\noindent\small
This paper incorporates works carried out during the first author's
visit to the IoA (Cambridge), 2011 September, which was in part
supported by the IoA's visitor grant.
JA is suppoted by the Chinese Academy of Sciences (CAS) Fellowships
for Young International Scientist, grant no.~2009Y2AJ7 and
the Nataional Natural Science Foundation of China (NSFC) Research Fund
for International Young Scientist.
AJD thanks the Science and Technology Facilities Council (STFC; UK)
for the award of a studentship.

%\newpage

%\vfill\bsp
\label{lastpage}

\begin{thebibliography}{}
\renewcommand\baselinestretch{0}
\footnotesize

\bibitem[An \& Evans(2011)]{AE11}
An J. H., Evans N. W., 2011, MNRAS, 413, 1744 (Paper I)

\bibitem[Anderson \& van der Marel(2010)]{AvdM10}
Anderson J., van~der~Marel R. P., 2010, ApJ, 710, 1032

\bibitem[Bahcall \& Tremaine(1981)]{BT81}
Bahcall J. N., Tremaine S., 1981, ApJ, 244, 805

\bibitem[Binney(1980)]{Bi80}
Binney J., 1980, MNRAS, 190, 873

\bibitem[Brown et al.(2005)]{Br05}
Brown W. R., %et al.,
Geller M. J., Kenyon S. J., Kurtz M. J., Allende~Prieto C., Beers T. C., Wilhelm R.,
2005, AJ, 130, 1097 

\bibitem[Crain et al.(2009)]{Ca09}
Crain R. A., et al., 2009, MNRAS, 399, 1773

\bibitem[Deason et al.(2011)]{De11}
Deason A. J., %et al.,
McCarthy I. G., Font A. S., Evans N. W., Frenk C. S., Belokurov V., Libeskind N. I., Crain R. A., Theuns T.,
2011, MNRAS, 415, 2607

\bibitem[Evans et al.(1997)Evans, H\"afner \& de Zeeuw]{Ev97}
Evans N. W., H\"afner R. M., de~Zeeuw P. T., 1997, MNRAS, 286, 315

\bibitem[Evans et al.(2003)]{Ev03}
Evans N. W., Wilkinson M. I., Perrett K. M., Bridges T. J., 2003, ApJ, 583, 752

\bibitem[Evans et al.(2011)Evans, An \& Deason]{Ev11}
Evans N. W., An J., Deason A. J., 2011, ApJ, 730, L26 (EAD)

\bibitem[Heisler et al.(1985)Heisler, Tremaine \& Bahcall]{He85}
Heisler J., Tremaine S., Bahcall J. N., 1985, ApJ, 298, 8

\bibitem[Leonard \& Merritt(1989)]{LM89}
Leonard P. J. T., Merritt D., 1989, ApJ, 339, 195

\bibitem[Limber \& Mathews(1960)]{Li60}
Limber D. N., Mathews W. G., 1960, ApJ, 132, 286

\bibitem[Lindegren et al.(2008)]{Li08}
Lindegren L., et al., 2008,
in Jin J., Platais I., Perryman M.A.C., eds, Proc.\ IAU Symposium No.~248,
A Giant Step: from Milli- to Micro-arcsecond Astrometry: Cambridge Univ.\ Press, Cambridge, p.217

\bibitem[Page(1952)]{Pa52}
Page T., 1952, ApJ, 116, 63 

\bibitem[Perryman et al.(2001)]{Pe01}
Perryman M. A. C., %et al.,
de~Boer K. S., Gilmore G., H{\o}g E., Lattanzi M. G., Lindegren L., Luri X., Mignard F., Pace O., de Zeeuw P. T.,
2001, A\&A, 369, 339

\bibitem[Perryman (2009)]{Pe09}
Perryman M., 2009,
Astronomical Applications of Astrometry: Ten Years of Exploitation of the {\it Hipparcos} Satellite Data. Cambridge Univ.\ Press, Cambridge

\bibitem[Piatek et al.(2005)]{Pi05}
Piatek S., %et al.,
Pryor C., Bristow P., Olszewski E. W., Harris H. C., Mateo M., Minniti D., Tinney C. G.,
2005, AJ, 130, 95

\bibitem[Piatek et al.(2006)]{Pi06}
Piatek S., %et al.,
Pryor C., Bristow P., Olszewski E. W., Harris H. C., Mateo M., Minniti D., Tinney C. G.,
2006, AJ, 131, 1445 

\bibitem[Scholz \& Irwin(1994)]{Sc94}
Scholz R. D., Irwin M. J., 1994,
in MacGillivray H.T., Thomson E.B., Lasker B.M., Reid I.N., Malin D.F., West R.M., Lorenz H., eds, Proc.\ IAU Symposium No.~161,
Astronomy from Wide-Field Imaging: Kluwer Academic Publ., Dordrecht, p.535

\bibitem[Springel et al.(2005)]{Sp05}
Springel V., et al., 2005, Nature, 435, 629

\bibitem[Trumpler \& Weaver(1953)]{SA}
Trumpler R. J., Weaver H. F., 1953,
Statistical Astronomy. Univ.\ of California Press, Berkeley
({\sc reprinted} 1962, Dover Publ., New York)

\bibitem[Turon et al.(2005)Turon, O'Flaherty, \& Perryman]{Tu05}
Turon C., O'Flaherty K. S., Perryman M. A. C., 2005,
Proc.\ the {\it Gaia} Symposium ``The Three-Dimensional Universe with {\it Gaia}''. European Space Agency Publ.

\bibitem[van der Marel \& Anderson(2010)]{vdMA10}
van~der~Marel R. P., Anderson J., 2010, ApJ, 710, 1063

\bibitem[Walker et al.(2008)Walker, Mateo, \& Olszewski]{Wa08}
Walker M. G., Mateo M., Olszewski E. W., 2008, ApJ, 688, L75 

\bibitem[Watkins et al.(2010)Watkins, Evans \& An]{WEA}
Watkins L. L., Evans N. W., An J. H., 2010, MNRAS, 406, 264 (WEA)

\bibitem[White(1981)]{Wh81}
White S. D. M., 1981, MNRAS, 195, 1037 

\bibitem[Wilkinson \& Evans(1999)]{WE99}
Wilkinson M. I., Evans N. W., 1999, MNRAS, 310, 645

\end{thebibliography}
\end{document}